\documentclass[12pt]{iopart}
\usepackage{iopams}
\usepackage{graphicx}

\begin{document}

\title[]{Spatial mode effects in a cavity EIT-based quantum memory with ion Coulomb crystals}

\author{Kasper R. Zangenberg, Aur\'elien Dantan, Michael Drewsen}

\address{QUANTOP, Danish National Research Foundation Center for Quantum Optics, Department of
Physics and Astronomy, Aarhus University, DK-8000 \AA rhus C., Denmark}
\begin{abstract}
Quantum storage and retrieval of light in ion Coulomb crystals using cavity electromagnetically
induced transparency is investigated theoretically. It is found that, when both the control and
probe fields are coupled to the same cavity mode, their transverse mode profile affects the
quantum memory efficiency in a non-trivial way. Under such conditions the control field parameters
and crystal dimensions that maximize the memory efficiency are calculated.
\end{abstract}

\submitto{\JPB}

\section{Introduction}

Motivated by applications in the field of quantum information processing
~\cite{duan01,lukin03,kimble08,hammerer10}, quantum memory devices are being investigated in a
variety of physical systems and with different techniques and protocols (for a review see
e.g.~\cite{lvovsky09,simon10}). Among the various criteria used to evaluate the performance of a
quantum memory, are generally of interest its fidelity, its efficiency, its storage time and its
multimode capacity~\cite{lvovsky09}. For optical quantum memories, in which an input light pulse is
stored into a material system and subsequently retrieved, the efficiency can simply be defined as
the ratio between the energies of the output and input pulses. For an important class of optical
quantum memories based on electromagnetically induced transparency (EIT) processes in atomic
ensembles~\cite{chaneliere05,eisaman05,simon07,appel08,honda08,cviklinski08,choi08,zhaob09,zhaor09},
the efficiency crucially depends on the optical depth of the ensemble~\cite{lvovsky09}. Enclosing
the atomic medium in an optical cavity allows for substantially increasing the effective optical
depth experienced by the light and cavity EIT protocols have been proposed to achieve high storage
efficiencies~\cite{lukin00,dantan04,zimmer06,dantan06,gorshkov07}.

Among the variety of atomic systems studied in connection with quantum memories, an ion Coulomb
crystals positioned in an optical cavity has been suggested as a good candidate to realize a
high-performance quantum memory, potentially meeting the criteria mentioned
above~\cite{mortensen05}. Recently, this analysis has been backed up by a series of key
experimental results. Substantial effective optical depths can indeed be realized by strongly
coupling large ion Coulomb crystals to a cavity field and long coherence times can be achieved in
such a system~\cite{herskind09,albert12}. Furthermore, strong coupling to various spatial cavity
modes has been demonstrated~\cite{dantan09}, which is promising for multimode storage. Finally,
cavity EIT has recently been observed with ion Coulomb crystals using an \textit{all}-cavity geometry in which both the probe
and control fields are coupled to the same cavity mode~\cite{albert11}.

In this specific cavity EIT geometry, it has been observed, both theoretically~\cite{dantan12} and
experimentally~\cite{albert11}, that the spatial transverse profile of the control field has a
significant effect on the probe transmission/reflection lineshapes and dynamics. This is in
contrast with more standard cavity EIT configurations in which the control field is
free-propagating and has typically a much larger extension than the probe
field~\cite{muller97,hernandez07,wu08,mucke10,kampschulte10,laupetre11,specht11}. Since the
all-cavity geometry is one natural realization of the scheme presented in~\cite{mortensen05}, it is
interesting to investigate its implications for the performance of such an ion Coulomb
crystal-based quantum memory. In this paper, we extend the existing theoretical models for
cavity EIT-based light storage and retrieval~\cite{lukin00,dantan04,zimmer06,dantan06,gorshkov07}
to the all-cavity configuration and numerically investigate the effect of the transverse mode
profile of the fields on the quantum memory efficiency. We find in particular that the optimal
efficiency depends not only on the cooperativity parameter, but also on the radial extension of the
crystal, as a result of the more complex spatial mode structure defined by the fields inside the
atomic medium. Using parameters taken from current experiments with ion Coulomb crystals~\cite{herskind09,albert11}, our
simulations predict that similarly high-efficiencies ($>90\%$) should however be obtainable in the
all-cavity configuration.

The paper is structured as follows: in sec.~\ref{sec:theoretical_model} the theoretical model for
the light storage and retrieval is presented, starting with a general description in
secs.~\ref{sec:description} and \ref{sec:equationsEIT}, an optimization of the quantum memory
efficiency in the standard configuration for smoothly-varying single-photon input pulses, following
the approach of ref.~\cite{gorshkov07}, in sec.~\ref{sec:optimization_extended} and the ``cylindrical
shell" model used for the simulations of the all-cavity configuration in
sec.~\ref{sec:effect_transverse}. Section~\ref{sec:numerical_results} presents the results of the
numerical simulations based on typical experimental parameters for ion Coulomb crystals
(sec.~\ref{sec:parameters}) and for light fields coupled to the cavity fundamental TEM$_{00}$ mode
(sec.~\ref{sec:resultsTEM00}) and a higher-order Laguerre-Gauss LG$_{01}$ mode
(sec.~\ref{sec:resultsLG01}). A brief conclusion on the implications of these results for experiments are given in
sec.~\ref{sec:conclusion}.

\section{Theoretical model}\label{sec:theoretical_model}

\subsection{Description of the model}\label{sec:description}
\begin{figure}[h]
\centering
\includegraphics[width=\columnwidth]{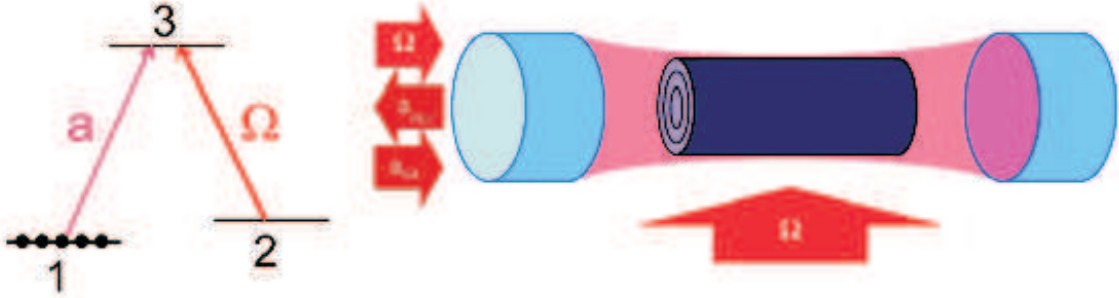}
\caption{(Color online) Left: Three-level $\Lambda$ atomic structure considered. Right: a
cylindrically symmetric atomic medium, composed of $\Lambda$ atoms, is enclosed in a single-ended
optical cavity where it interacts with a probe field and a control field in an EIT situation. Two
configurations are investigated: one in which both field are coupled to the same cavity mode - and
have therefore the same waist, and one in which the control field has a constant intensity profile
over the ensemble.} \label{fig:fig1}
\end{figure}

Inspired by the experiments of~\cite{albert11} we base our description of the light-matter
interaction on the model developed in~\cite{dantan12}. We consider an ensemble of three-level
$\Lambda$ atoms with two ground states, $|1\rangle$ and $|2\rangle$, and an excited state
$|3\rangle$. The atoms interact with a cavity probe field on the $|1\rangle\longrightarrow
|3\rangle$ transition and a classical control field $\Omega$ on the $|2\rangle\longrightarrow
|3\rangle$ transition. Both fields are assumed to be resonant with the atomic transitions and the
cavity is tuned to resonance with the probe field. Since we are mostly interested in discussing the
effects of the transverse mode profile of the fields, we neglect the longitudinal variation of
the atom-field couplings. This goes for atoms in a running-wave
cavity~\cite{wu08,laupetre11,nagorny03}, but can also apply to a standing-wave cavity geometry for
atoms with either well-defined positions with respect to the cavity
standing-wave~\cite{hernandez07,brennecke07,colombe07} or atoms whose motion average out the longitudinal
standing-wave structure of the fields over the relevant timescales~\cite{albert11,dantan12}. The atomic ensemble
extension is also assumed to be much smaller than the cavity field Rayleigh range. The cavity is
furthermore taken to be single-ended and with lossless mirrors (fig.~\ref{fig:fig1}).

We will also assume that the fields injected into the cavity are \textit{smooth} pulses whose
envelopes are slowly varying with respect to the cavity field and atomic decay rates, in the sense
of~\cite{gorshkov07}. For the light storage and retrieval dynamics we will follow the approach of
refs.~\cite{lukin00,dantan06,gorshkov07} and assume a standard write-store-read temporal sequence.
During the write phase the input probe field is injected into the cavity and the control field is
adiabatically turned off to ensure \textit{temporal impedance matching}~\cite{zimmer06}. Both
fields are turned off during the store phase, and the control field is adiabatically turned back on
during the readout, causing the emission of an output probe pulse.

Denoting by $\hat{\sigma}_{\mu\nu}^{(k)}=|\mu\rangle\langle\nu|_k$ the individual atomic operator
for the $k-$th atom positioned at $\textbf{r}_k$ and by $\hat{a}$ the annihilation operator for the
intracavity probe field, the interaction Hamiltonian in the rotating wave approximation and the
rotating frame reads \begin{eqnarray}H=-\hbar
g\sum_{k}^{}\Psi_p(r_k)\hat{\sigma}_{31}^{(k)}\hat{a}-\hbar\Omega(t)\sum_{k}^{}\Psi_c(r_k)\hat{\sigma}_{32}^{(k)}+\textrm{h.c.}\end{eqnarray}
where $g$ is the single-atom maximal coupling rate (at the center of the cavity mode),
$\Psi_{p,c}(r)$ are the probe and control field transverse mode profiles (with the longitudinal
dependence neglected) and $\Omega(t)$ the time-varying Rabi frequency of the control field.

\subsection{Equations of motion in EIT}\label{sec:equationsEIT}
We consider a typical EIT regime in which all the atoms are initially in level $|1\rangle$ and in which
the control field is much more intense than the probe field, assumed to be at the single-photon
level. One can then assume that almost all the atoms stay in $|1\rangle$ at all times and perform a
standard first-order treatment in the probe field~\cite{lukin03}. The Heisenberg equations of motion for the
relevant operators, namely the intracavity probe field $\hat{a}$, the atomic optical coherences $\hat{\sigma}_{13}^{(k)}$ and the atomic ground state
coherences $\hat{\sigma}_{12}^{(k)}$, are given by
\begin{eqnarray}
\label{eq:a_quantum}\dot{\hat{a}}&=-\kappa \hat{a}+ig\sum_{k}^{}\Psi_p(r_k)\hat{\sigma}_{13}^{(k)}+\sqrt{2\kappa}\hat{a}_{in},\\
\label{eq:sigma13_quantum}\dot{\hat{\sigma}}_{13}^{(k)}&=-\gamma\hat{\sigma}_{13}^{(k)}+ig\Psi_p(r_k)\hat{a}+i\Omega(t)\Psi_c(r_k)\hat{\sigma}_{12}^{(k)}+\hat{F}_{13}^{(k)},\\
\label{eq:12_quantum}\dot{\hat{\sigma}}_{12}^{(k)}&=-\gamma_0\hat{\sigma}_{12}^{(k)}+i\Omega^*(t)\Psi_c(r_k)\hat{\sigma}_{13}^{(k)}+\hat{F}_{13}^{(k)},
\end{eqnarray}
and the input-output relation
\begin{eqnarray}
\label{eq:a_out_quantum}\hat{a}_{out}=\sqrt{2\kappa}\hat{a}-\hat{a}_{in},
\end{eqnarray}
where $\kappa$ is the cavity field decay rate, $\hat{a}_{in}$ and $\hat{a}_{out}$ are the annihilation operators associated with the input and output probe fields, respectively.
$\gamma$ and $\gamma_0$ are the atomic dipole and ground state coherence decay rates, respectively,
and $\hat{F}_{13}^{(k)}$ and $\hat{F}_{12}^{(k)}$ the corresponding Langevin noise operators.

Following ref.~\cite{gorshkov07}, we assume a single-photon input and calculate the quantum memory
efficiency by first solving in time the semiclassical counterparts of~\eref{eq:a_quantum}-\eref{eq:a_out_quantum} for given input probe and control field pulses
\begin{eqnarray}
\dot{a}&=-\kappa a+ig\sum_{k}^{}\Psi_p(r_k)\sigma_{13}^{(k)}+\sqrt{2\kappa}a_{in},\label{eq:a_complex}\\
\dot{\sigma}_{13}^{(k)}&=-\gamma\sigma_{13}^{(k)}+ig\Psi_p(r_k)a+i\Omega(t)\Psi_c(r_k)\sigma_{12}^{(k)},\\
\dot{\sigma}_{12}^{(k)}&=-\gamma_0\sigma_{12}^{(k)}+i\Omega^*(t)\Psi_c(r_k)\sigma_{13}^{(k)},\\
a_{out}&=\sqrt{2\kappa}a-a_{in},\label{eq:a_out_complex}
\end{eqnarray}
and, secondly, by computing
\begin{equation}\eta_{tot}\equiv\frac{\int_{r}|a_{out}(t)|^2dt}{\int_{w}|a_{in}(t)|^2dt},\end{equation}
where the subscripts $w$ and $r$ refer to a summation over the whole duration of the write and read process, respectively. $\eta_{tot}$ then represents the ratio of the number of retrieved photons to the number of incoming photons, and provides a good measure of the quality of the mapping (for other measures see e.g.~\cite{lvovsky09,simon10}). To simplify the discussion we will in the following neglect the decay of the ground state coherence during the whole process and set $\gamma_0=0$. The duration of the storage phase is then simply chosen such that the dynamics of the write and read phases occur in well-separated time windows.

\subsection{Efficiency optimization in the case of an extended control field}\label{sec:optimization_extended}

If the waist of the control field is much larger than the probe field, as in many EIT experiments,
one can neglect the transverse variations of the control field intensity over the section of the
atomic ensemble that interacts with the probe field. Following, e.g.~\cite{dantan12}, one can
define an effective number of atoms interacting with the probe field
\begin{equation}\label{eq:N} N=\sum_{k}\Psi_p(r_k)^2\end{equation}
and collective operators for the ground state coherence and the optical dipole by
\begin{eqnarray}
\hat{S}=\frac{1}{\sqrt{N}}\sum_{k}^{}\Psi_p(r_k)\hat{\sigma}_{12}^{(k)},\hspace{0.2cm}\hat{P}=\frac{1}{\sqrt{N}}\sum_{k}^{}\Psi_p(r_k)\hat{\sigma}_{13}^{(k)}.
\end{eqnarray}
Equations~\eref{eq:a_complex}-\eref{eq:a_out_complex} can be straightforwardly rewritten as
\begin{eqnarray}
\label{eq:a_extended}\dot{a}&=-\kappa a+ig_NP+\sqrt{2\kappa}a_{in},\\
\label{eq:P_extended}\dot{P}&=-\gamma P+ig_Na+i\Omega(t)S,\\
\label{eq:S_extended}\dot{S}&=i\Omega^*(t)P,\\
a_{out}&=\sqrt{2\kappa}a-a_{in},
\end{eqnarray} where $g_N=g\sqrt{N}$ is the collectively enhanced coupling rate~\cite{lukin03}.

In this collective mode picture, the input photonic state is mapped during the write phase onto a collective spin-wave described by $S$ and one can define a \textit{write} efficiency by taking the ratio of the number of atomic excitations and the number of input photons,
\begin{equation}\eta_w=\frac{|S(T_w)|^2}{\int_w|a_{in}(t)|^2dt},\end{equation}
where $T_w$ is the end time of the write phase~\cite{gorshkov07}. Similarly, one can define
a readout efficiency by the ratio of the number of output photons and the number of
atomic excitations before readout
\begin{equation}\eta_r=\frac{\int_w|a_{out}(t)|^2dt}{|S(T_r)|^2},\end{equation} where $T_r$ is the
start time of the read phase. In the adiabatic limit, i.e. for input pulses with duration $T$ such
that $2TC\gamma\gg 1$, one can derive the optimal control pulse that maximizes the read and write
efficiencies, which can be shown to scale as~\cite{dantan06,gorshkov07}
\begin{equation}\eta_{w,r}^{opt}=\frac{2C}{1+2C},\end{equation} where
\begin{equation} C=\frac{g^2N}{2\kappa\gamma}\end{equation}
is the cooperativity parameter~\footnote{Note the factor 2 difference with respect to~\cite{gorshkov07}.}. In absence of decoherence during the storage phase, the optimal total efficiency thus scales as \begin{equation}\label{eq:eta_tot_opt}\eta_{tot}^{opt}=\left(\frac{2C}{1+2C}\right)^2,\end{equation}
and increases with the effective number of atoms defined by the spatial overlap of the ensemble and the probe field [Eq.~\eref{eq:N}].

\subsection{Effect of the control field's transverse profile}\label{sec:effect_transverse}

If the waist of the control field $w_c$ is no longer very large, but comparable to that of the
probe field $w_p$ -- as it will be the case in an all-cavity geometry for
instance~\cite{albert11,dantan12} -- the previous results no longer apply, and one must evaluate
the effect of the control field transverse profile on the storage and retrieval efficiencies.
With the approximations made in Sec.~\ref{sec:equationsEIT} and having specifically in mind ion
Coulomb crystals as the physical storage medium, we assimilate the atomic ensemble to a cylinder
with length $L$ and radius $R$ (fig.~\ref{fig:fig1}), and slice it into $n$ cylindrical shells of
thickness $d\ll w_p,w_c$ ($R=nd$). We also assume that $d$ is much larger than the mean
interparticle distance. Although it is not essential, we take the atomic density $\rho$ to be
constant throughout the ensemble, which is the case for large ion Coulomb crystals in linear Paul
traps~\cite{hornekaer01}, and we consider cavity modes with cylindrical symmetry. We proceed by
defining collective operators for the $j$th-slice as
\begin{equation} \hat{P}_j=\sigma n_j \hat{\sigma}_{13}^{(j)},\hspace{0.2cm}\hat{S}_j=\sigma n_j \hat{\sigma}_{12}^{(j)},\end{equation}
where the subscript $j$ refers to an atom in the $j$-th slice with position $r_j=d(j-1/2)$
($j=1..n$) and where $\sigma=\rho L$ is the atomic cross-sectional density. The corresponding
semiclassical equations of motion are then
\begin{eqnarray}
\label{eq:a_transverse}\dot{a}&=-\kappa a+ig\sum_{j=1}^{n}\Psi_p(r_j)P_j+\sqrt{2\kappa}a_{in},\\
\label{eq:P_j_transverse}\dot{P}_{j}&=-\gamma P_{j}+ig\sigma n_j\Psi_p(r_j)a+i\Omega(t)\Psi_c(r_j)S_{j},\\
\label{eq:S_j_transverse}\dot{S}_{j}&=i\Omega^*(t)\Psi_c(r_j)S_j,
\end{eqnarray} where the modefunctions $\Psi_{p,c}$ are evaluated at $r_j$.

It is clear from eqs.~\eref{eq:a_transverse},\eref{eq:P_j_transverse},\eref{eq:S_j_transverse} that,
unless $|\Psi_c(r_j)|=1$ like in the previous section, it is no longer possible to define
collective spatial eigenmodes of the problem that would yield closed equations of the form
\eref{eq:a_extended},\eref{eq:P_extended},\eref{eq:S_extended}. In particular, these equations show
that adjacent shells are coupled together by the control field. The spatial mapping of the probe
field onto the ground state spin now depends on the control field transverse profile, in addition
to that of the probe field. Because of the intershell coupling during the mapping the radial
extension of the ensemble now becomes a parameter which affects the memory efficiency in a
non-trivial way.

In the following section we numerically solve these equations of motion for a fixed probe field
pulse basing ourselves on the analytical control field pulse derived from the temporal optimization
of sec.~\ref{sec:optimization_extended} in absence of effects due to the control field transverse
profile. Note that, because of the impossibility of defining analytically a spatial collective
spin-mode during the write or read phase, and thereby of defining write or read efficiencies, this optimization
is performed numerically using the total efficiency $\eta_{tot}$ as a figure of merit.

\section{Numerical results}\label{sec:numerical_results}

\subsection{Physical system considered and input parameters}\label{sec:parameters}
To solve the problem of optimizing the quantum memory efficiency in the conditions of the previous
section we take for the physical storage medium an ion Coulomb crystal, trapped and laser-cooled in
a linear Paul trap, with the optical cavity positioned along the trap axis, as
in~\cite{herskind09,dantan09,albert12,albert11}. Although single-component ion Coulomb
crystals (i.e. consisting of only ion species) have spheroidal shape and may therefore deviate from
the cylindrical shell model (unless they are sufficiently prolate), the inner component of a
prolate two-species crystal can be assimilated to a good approximation to a uniform density
cylinder~\cite{hornekaer01,herskind08}. We consider the cavity EIT configuration used in~\cite{albert11},
in which both the control and probe fields are frequency degenerate and orthogonally polarized in
order to create EIT between Zeeman sub-states of the $3d\; ^3D_{3/2}$ sub-level of $^{40}$Ca$^+$. In
these experiments both fields are coupled resonantly or near-resonantly to the same cavity mode
(TEM$_{00}$). We thus take Gaussian transverse profiles $\Psi_{p,c}(r)=\exp(-r^2/w_{p,c}^2)$, and
compare the extended control field configuration ($w_c\rightarrow\infty$) and the finite control field waist configuration ($w_c=w_p$).

For the 11.8 mm-long, close to confocal cavity of~\cite{albert11} with an incoupling mirror
transmission of 1500 ppm and an interaction on the $3d\; ^3D_{3/2},m_J=+3/2\rightarrow 4p\;
^2P_{1/2},m_J=+1/2$ (probe) and $3d\; ^3D_{3/2},m_J=-1/2\rightarrow 4p\; ^2P_{1/2},m_J=+1/2$
(control) transitions, one finds ($g,\kappa,\gamma)=2\pi\times(0.37,1.5,11.3)$ MHz~\footnote{Note that, in constrast to ~\cite{herskind09,albert11}, the single-ion coupling rate has been scaled by a factor $1/\sqrt{2}$ to account for the longitudinal averaging over the standing-wave structure.}. With a radius
of curvature of 10 mm, the waist of the probe field at the center of the cavity is $w_p=37$ $\mu$m.
For crystals with typical length of a few mm and radius of up to a few hundreds of microns,
neglecting the longitudinal curvature of the fields over the crystal modevolume is
well-justified~\cite{herskind09,dantan09}.

For the sake of the discussion, we assume for the probe field a hyperbolic secant input pulse of the form
\begin{equation}
a_{in}(t)=\frac{1}{\sqrt{T}}\textrm{sech}(2t/T),\label{eq:a_in(t)}
\end{equation} where $T$ is the probe pulse duration~\footnote{This form for the probe pulse is taken for convenience, as one gets an analytical expression for the optimal control field pulse~\cite{lukin00}, but the numerical simulations show that the exact form of the input pulse is not critical.}.
In the adiabatic limit ($TC\gamma\gg 1)$ considered previously in
sec.~\ref{sec:optimization_extended} and the extended control field configuration
($w_c\rightarrow\infty$), the control field pulses which optimize the write and read efficiencies
are given by~\cite{gorshkov07}
\begin{eqnarray}
\Omega_w (t)=A\sqrt{\frac{2\gamma(1+2C)}{T}}\frac{1}{\sqrt{1+\exp(4t/T)}}\label{eq:Omega_w(t)}
\end{eqnarray}
for the write phase, and its time-reversed counterpart \begin{equation}\Omega_r (t)=\Omega_w(-t+T_r+T_s),\label{eq:Omega_r(t)}\end{equation} for the readout phase. In the extended configuration the prefactor $A$ is equal to unity. As we will see in the next section, the previous control field temporal profiles are still found to be
optimal with respect to maximizing the efficiency in the finite waist configuration, the main difference being in the optimal control field amplitude scaling factor $A$.

\subsection{Results for the TEM$_{00}$ mode}\label{sec:resultsTEM00}

\begin{figure}[h]
\centering
\includegraphics[width=0.55\columnwidth]{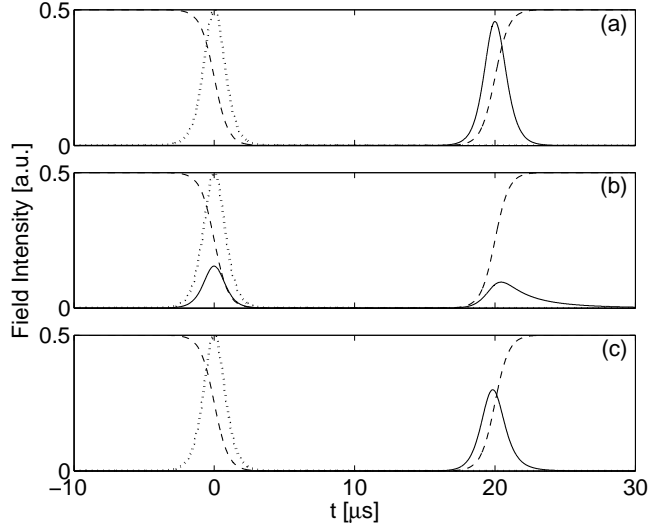}
\caption{Temporal storage and retrieval sequence for (a) an extended control field ($w_c\rightarrow\infty$),
(b) a control field with finite waist ($w_c=w_p$) and non-optimized amplitude ($A=1$) and (c) a control
field with finite waist ($w_c=w_p$) and optimized amplitude ($A=2.45$). The dotted and solid curves are the normalized input and output field intensities, $|a_{in}(t)|^2$ and $|a_{out}(t)|^2$, respectively. The dashed curve shows the control field intensity $\Omega(t)^2$, scaled such that its maximum value is 0.5. $T_r=T_w=10$ $\mu$s. See text for the value of the other parameters.} \label{fig:temporal_sequence}
\end{figure}

Figure~\ref{fig:temporal_sequence}a shows the results of a storage and retrieval sequence for a
crystal with density $\rho=6.1\times 10^8$ cm$^{-3}$, length $L=3$ mm and radius $R=100$ $\mu$m, an
input probe pulse of the form \eref{eq:a_in(t)} with duration $T=2$ $\mu$s and an extended control
field of the form \eref{eq:Omega_w(t)},\eref{eq:Omega_r(t)} with $A=1$. For such a large crystal
($R\gg w_c$), the effective number of ions as defined by \eref{eq:N} is $N=3936$, yielding a
cooperativity parameter $C\simeq 16.7$. The write and read efficiencies are found to be
$\eta_w=0.970$ and $\eta_r=0.971$, respectively, close to the theoretical value of
$\eta_{r,w}^{opt}=0.971$, and yielding a total efficiency $\eta_{tot}=0.942$
($\eta_{tot}^{opt}=0.943$). Figure~\ref{fig:temporal_sequence}b shows the results of the same
sequence and parameters for a control field with $w_c=w_p$. It is clear that, during the write
phase, perfect temporal impedance matching is not achieved, as a substantial amount of the incoming
light is reflected. This is in itself not surprising as the prefactor $A=1$ is only optimal in the
extended configuration, and one expects that, in the finite waist configuration, the ions see on
average a control field with lower Rabi frequency. As can be seen from
fig.~\ref{fig:temporal_sequence}c, close to perfect impedance matching can be recovered by
increasing the control field amplitude ($A\simeq 2.45$). The total efficiency $\eta_{tot}=0.667$
remains, however, lower than in the corresponding extended configuration. We checked numerically that
varying the control field pulse switching time and shape, or having different amplitudes/time
evolutions profiles during write and read, does not increase the efficiency. We also checked that
these results do not significantly depend on $T$ as long as one stays in the adiabatic limit.

\begin{figure}[h]
\centering
\includegraphics[width=0.55\columnwidth]{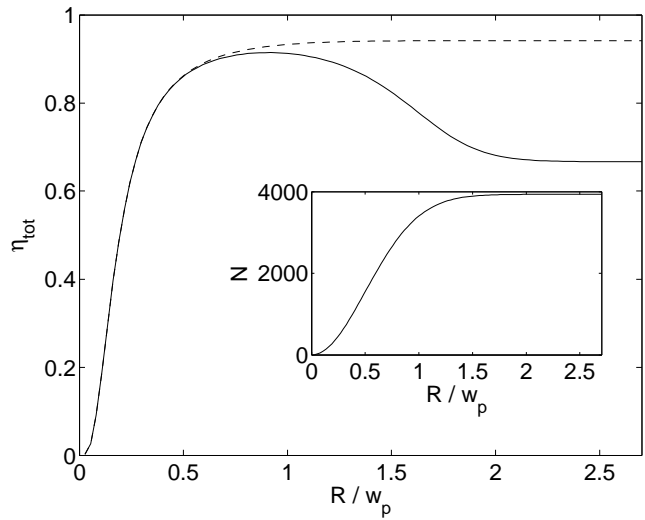}
\caption{Variation of the total efficiency $\eta_{tot}$ as a function of the crystal
radius $R$, for the same parameters as in fig.~\ref{fig:temporal_sequence}. The dashed line shows
the extended control field case while the solid line shows the finite waist case for which the
control field amplitude has been optimized for each radius. The inset shows the variation of the
effective number of ions [eq.~\eref{eq:N}] as a function of the crystal radius $R$.}
\label{fig:variation_R}
\end{figure}

The lack of critical dependence of the optimal efficiency with respect to the temporal parameters
found in the simulations seems to indicate, in agreement with the analysis of
sec.~\ref{sec:effect_transverse}, that the spatial profile of the control field now plays a
significant role in the mapping process. To investigate this effect further we show in
fig.~\ref{fig:variation_R} the variation of the effective number of ions and the (temporally)
optimized efficiency as a function of the crystal radius in the two configurations. The other
parameters are kept the same as previously. While in the extended control field configuration the
efficiency increases, together with $N$, with the crystal radius and saturates when $R\gg w_p$, it
reaches a maximum for $R\sim w_p$ in the finite waist configuration, before decreasing and reaching
a constant level at high radius. The total efficiency peaks at $R\simeq 0.95w_p$ with a value of
0.914 ($A=1.5$), closer to the theoretical value of $\eta_{tot}^{opt}=0.932$ (for this radius $N=3279$ and
$C=13.9$). The decrease for $R>w_p$ radius may appear surprising, since more ions are being added to the crystal
and one could expect an enhanced efficiency due to the stronger coupling to the probe field.
However, for a finite waist control field, the spatial spin mode defined during writing -- or, for
that matter, during reading -- is no longer that of the probe, but depends on the overlap of both
the probe \textit{and} the control field transverse profiles in the crystal. As the radius of the
crystal is increased the stored photonic excitation is spread more and more over shells with higher
radius. The spatial atomic mode then no longer resembles the spatial mode of the probe, which leads
to a decrease in efficiency in the writing process. A similar phenomenon then takes place in the
reading process, as the atomic spatial excitation profile is no longer optimally matched to the spatial
light mode profile.

\begin{figure}[h]
\centering
\includegraphics[width=0.55\columnwidth]{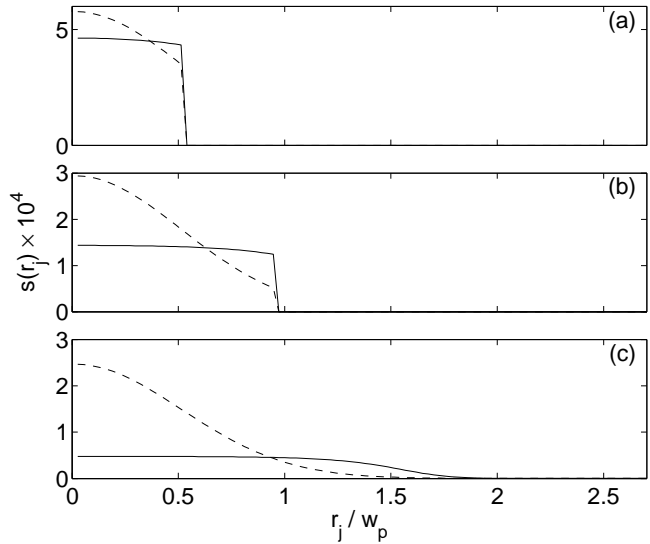}
\caption{Variation of the radial density of excitations after writing in the the
extended (dashed) and finite waist (solid) configurations, for different crystal radii [(a)
$R=0.5w_p$ (b) $R=0.95w_p$ (c) $R=2.7w_p$]. Other parameters as in Fig.~\ref{fig:variation_R}.}
\label{fig:variation_radial_proba}
\end{figure}

To illustrate this phenomenon, we show in fig.~\ref{fig:variation_radial_proba} the variation of
the radial density of excitations after writing, which is proportional to the surface probability
of finding the photonic excitation in the $j$-th shell, at the end of the write phase, again in both
configurations and for crystals with different radii. The radial density of excitations $s(r_j)$ is
obtained by normalizing the squared modulus of the $j$-th shell operator mean value $S_j(T_w)$ by
the number of ions in the shell $n_j$,
\begin{equation}
s(r_j)=|S_j(T_w)|/n_j|^2
\end{equation}
In the extended configuration, one sees that $s(r_j)$ reproduces well the spatial Gaussian mode
profile of the probe field, as expected from the analytical predictions of
sec.~\ref{sec:optimization_extended}. In the finite waist configuration, the spatial atomic mode
defined by $s(r_j)$ has reasonable overlap with the ideal extended configuration mode (i.e. the
probe field mode) for $R\lesssim w_p$, but clearly deviates from it as the crystal radius increases
and the coupling between the shells causes the excitation to spread more and more radially into the
crystal.

\begin{figure}[h]
\centering
\includegraphics[width=0.55\columnwidth]{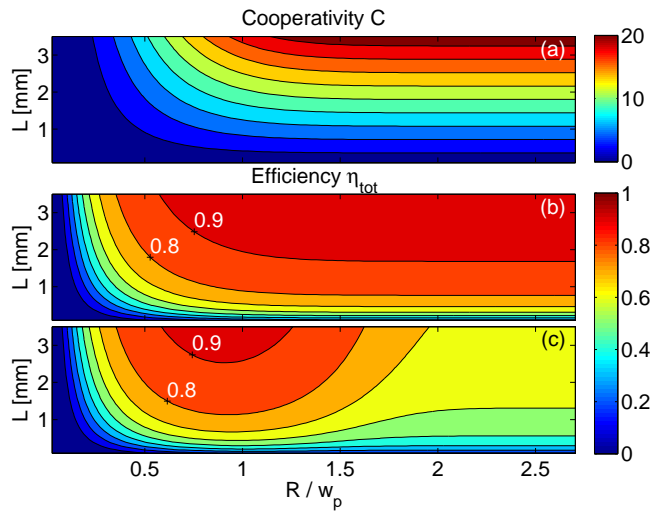}
\caption{(Color online) (a) Variation of the cooperativity as a function of the crystal dimensions
$L$ and $R$. (b) Variation of the optimized efficiency in the extended configuration versus $L$ and
$R$. (c) Variation of the optimized efficiency in the finite control field waist configuration
($w_c=w_p$) versus $L$ and $R$. The crystal density is $\rho=6.1\times 10^8$ cm$^{-3}$ and the pulse
duration $T=2$ $\mu$s.} \label{fig:summary}
\end{figure}

To summarize the results and show in particular that the decrease in efficiency for large radii is
always substantial in the finite waist configuration, regardless of the length or density of the
crystal, fig.~\ref{fig:summary} shows the variation of the optimized efficiency as a function of
the crystal dimensions, $L$ and $R$, for a density of $6.1\times 10^8$ cm$^{-3}$. The range chosen
for the dimensions is typical of current experiments with ion Coulomb crystals in
cavity~\cite{herskind09,herskind08}. The optimized quantum memory efficiency in the extended
control field configuration is found to agree well with the predictions from the analytical model
[eq.~\ref{eq:eta_tot_opt}], and is an increasing function of the crystal length for all radii. This
is no longer true in the finite waist configuration for which an optimal radius - of the order of
$w_p$ - exists, for all lengths. However, it can be seen that by choosing the radius of the
crystal appropriately to ``match" the waist of the cavity mode one can achieve similarly high quantum memory
efficiencies ($>90 \%$) as in the extended configuration.

\subsection{Higher-order modes}\label{sec:resultsLG01}

In this last section we investigate these spatial mode effects on the storage using
higher-order spatial cavity modes. On the one hand, this is motivated by the fact that collective strong
coupling with higher-order cavity modes has been demonstrated using ion Coulomb
crystals~\cite{dantan09}, which is promising for multimode (spatial) storage. On the other hand, in
view of the previous results, one can wonder how the conclusions drawn for the TEM$_{00}$ mode
generally hold for higher-order cavity modes, and in particular, if some modes are less sensitive
to these spatial effects. A general analysis is beyond the scope of the present paper and we will
only focus in this last section on the case of a first-order Laguerre-Gauss cavity mode LG$_{01}$,
which preserves the cylindrical symmetry of the problem. We thus assume that the probe field radial
modefunction is now given by $\Psi_p(r)=\sqrt{2}(r/w_p)\exp(-r^2/w_p^2)$. We then compare the
quantum memory efficiency in an extended control field configuration and in a configuration where
the control field has the same transverse profile as the probe field ($\Psi_c(r)=\Psi_p(r)$).

\begin{figure}[h]
\centering
\includegraphics[width=0.55\columnwidth]{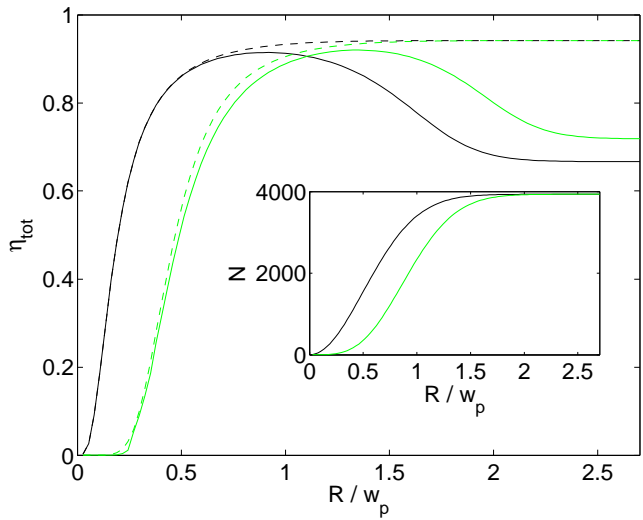}
\caption{(Color online) Variation of the quantum memory efficiency $\eta_{opt}$ as a function of
the crystal radius $R$, for the TEM$_{00}$ (black) and LG$_{01}$ (green) modes and for a crystal
with $L=3$ mm and $\rho=6\times 10^8$ cm$^{-3}$. The dashed lines show the extended configuration
and the solid lines the finite waist configuration. The inset shows the corresponding variations of
the effective number of ions [eq.~\eref{eq:N}] as a function of $R$, for both modes.}
\label{fig:variation_R_laguerre}
\end{figure}

\begin{figure}[h]
\centering
\includegraphics[width=0.55\columnwidth]{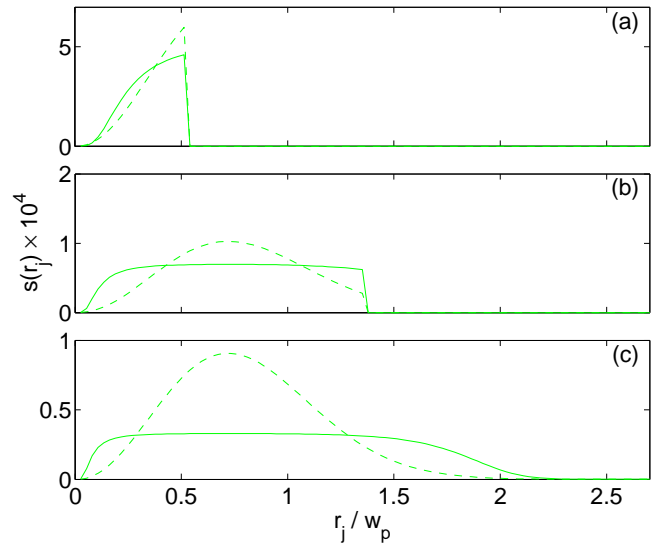}
\caption{(Color online) Variation of the mean surface number of excitations after writing in the
the extended (dashed) and finite control field waist (solid) configurations, for different crystal
radii [(a) $R=0.5w_p$ (b) $R=1.35w_p$ (c) $R=2.7w_p$] and for the LG$_{01}$ mode. Other parameters as in
Fig.~\ref{fig:variation_R_laguerre}.} \label{fig:variation_radial_proba_laguerre}
\end{figure}

Figure~\ref{fig:variation_R_laguerre}a shows the variation of the effective number of ions defined
by eq.~\eref{eq:N} as a function of the crystal radius, for a crystal with fixed length and density
($L=3$ mm and $\rho=6.1\times 10^8$ cm$^{-3}$). As expected, $N$ increases less rapidly at small
radii than for the fundamental mode, on account of the lower coupling at the center of the mode,
but saturates at the same value for large radii, because the orthonormal character of the
modefunctions. In fig.~\ref{fig:variation_R_laguerre}b are represented the corresponding variations
of the quantum memory efficiency, both for the TEM$_{00}$ and LG$_{01}$ modes and in the extended
and finite waist configurations. If a similar behavior is qualitatively observed for the LG$_{01}$
mode, one can see that the optimal radius is larger for the LG$_{01}$ mode ($R\simeq 1.35 w_p$)
than for the TEM$_{00}$ mode ($R\simeq 0.95 w_p$). This can be explained by the fact that the
regions of high radial intensity for the fields are now located farther away from $r=0$, so that
the spreading of the photonic excitations into large radius shells occurs at larger $R$ and is less
pronounced. One can carry out a similar analysis as for the fundamental mode and calculate the
variation of the radial density of excitations after writing as a function of the crystal radius.
The results are represented in fig.~\ref{fig:variation_radial_proba_laguerre} and show the same
qualitative conclusions as drawn previously in the case of the fundamental mode.

\section{Conclusion}\label{sec:conclusion}

We have investigated the efficiency of a cavity EIT-based quantum memory in which both control and
probe fields are coupled to the same cavity mode. Due to the complex spatial atomic mode defined
during the EIT process between both fields during the write and read phase of the memory, the
optimal efficiency is found to depend not only on the cooperativity parameter, but also the crystal
radius. Using parameters from current experiments with ion Coulomb
crystals~\cite{herskind09,albert11}, our simulations predict that high-efficiencies ($>90\%$)
should however be obtainable in this specific configuration.

This theoretical investigation clearly implies that an experimental realization of such an
all-cavity EIT quantum memory based on a cylindrical ion Coulomb crystal can be optimized by
choosing a crystal with radius matching the waist of the cavity modes. For single species crystals
as used in~\cite{herskind09,albert11}, the optimum crystal radius has been found to differ slightly because of
the spheroidal (non-cylindrical) shape of such crystals~\cite{hornekaer01}, but similar conclusions hold. For
two-species crystals, the lighter species typically takes the shape of a nearly perfect cylindrical
rod surrounded by the other ion species~\cite{hornekaer01,herskind08,mortensen07}, and matches
perfectly the situation considered theoretically.  Applying isotope selective
photoionization~\cite{mortensen04}, Coulomb crystals consisting of two calcium isotopes can easily
be created with varying compositions and shapes~\cite{herskind08,mortensen07}, providing an ideal
situation to test the predictions and optimize the storage conditions.

\ack We acknowledge financial support from the Carlsberg Foundation and the EU via the FP7 projects 'Physics of Ion Coulomb Crystals' (PICC) and 'Circuit and Cavity Quantum Electrodynamics' (CCQED), as well as useful discussions with Ian D. Leroux and Rasmus B. Linnet.


\section*{References}
\begin{thebibliography}{100}


\bibitem{duan01} Duan L M, Lukin M D, Cirac J I and Zoller P 2001 {\it Nature} {\bf 414} 413

\bibitem{lukin03} Lukin M D 2003 {\it Rev. Mod. Phys.} {\bf 75} 457

\bibitem{kimble08} Kimble H J 2008 {\it Nature} {\bf 453} 1023

\bibitem{hammerer10} Hammerer K, S{\o}rensen A S and Polzik E S 2010 {\it Rev. Mod. Phys.} {\bf 82} 1041

\bibitem{lvovsky09} Lvovksy A I, Sanders B and Tittel W 2009 {\it Nature Photon.} {\bf 706}

\bibitem{simon10} Simon C \textit{et al.} 2010 {\it Eur. Phys. J. D} {\bf 58} 1


\bibitem{chaneliere05} Chaneli\`{e}re T, Matsukevich D N, Jenkins S D, Lan S-Y, Kennedy T A B and Kuzmich A 2005 {\it Nature} {\bf 438} 833

\bibitem{eisaman05} Eisaman M D, Andr\'{e} A, Massou F, Fleischhauer M, Zibrov A S and Lukin M D 2005 {\it Nature} {\bf 438} 837

\bibitem{simon07} Simon J, Tanji H, Ghosh S and Vuletic V 2007 {\it Nature Phys.} {\bf 3} 765

\bibitem{appel08} Appel J, Figueroa E, Korystov D, Lobino M and Lvovsky A 2008 {\it Phys. Rev. Lett.} {\bf 100} 093602

\bibitem{honda08} Honda K, Akamatsu D, Arikawa M, Yokoi Y, Akiba K, Nagatsuka S, Tanimura T, Furusawa A and Kozuma M 2008 {\it Phys. Rev. Lett.} {\bf 100} 093601

\bibitem{cviklinski08} Cviklinski J, Ortalo J, Laurat J, Bramati A, Pinard M and Giacobino E 2008 {\it Phys. Rev. Lett.} {\bf 101} 133601

\bibitem{choi08} Choi K S, Deng H, Laurat J and Kimble H J 2008 {\it Nature} {\bf 452} 67

\bibitem{zhaob09} Zhao B, Chen Y-A, Bao X-H, Strassel T, Chu C-S, Jin X-M, Schmiedmayer J, Yuan Z-S, Chen S and Pan J-W 2009 {\it Nature Phys.} {\bf 5} 95

\bibitem{zhaor09} Zhao R, Dudin Y O, Jenkins S D, Campbell C J, Matsukevich D N, Kennedy T A B and Kuzmich A 2009 {\it Nature Phys.} {\bf 5} 100


\bibitem{lukin00} Lukin M D, Yelin S F and Fleischhauer M 200 {\it Phys. Rev. Lett.} {\bf 84} 4232

\bibitem{dantan04} Dantan A and Pinard M 2004 {\it Phys. Rev. A} {\bf 69} 043810

\bibitem{zimmer06} Zimmer F E, Andr\'{e} A, Lukin M D and Fleischhauer M 2006 {\it Opt. Comm.} {\bf 264} 441

\bibitem{dantan06} Dantan A, Cviklinski J, Pinard M and Grangier P 2006 {\it Phys. Rev. A} {\bf 73} 032338

\bibitem{gorshkov07} Gorshkov A, Andr\'{e} A, Lukin M D and S{\o}rensen A S 2007 {\it Phys. Rev. A} {\bf 76} 033804


\bibitem{mortensen05} Mortensen A, PhD thesis, Aarhus University (2005).

\bibitem{herskind09} Herskind P F, Dantan A, Marler J P, Albert M and Drewsen M 2009 {\it Nature Phys.} \textbf{5} 494

\bibitem{albert12} Albert M, Marler J P, Herskind P F, Dantan A and Drewsen M 2012 {\it Phys. Rev. A} in press (\textit{Preprint} arxiv:1108.0528)

\bibitem{dantan09} Dantan A, Albert M, Marler J P, Herskind P F and Drewsen M 2009 {\it Phys. Rev. A} {\bf 80} 041802(R)

\bibitem{albert11} Albert M, Dantan A and Drewsen M 2011 {\it Nature Photon.} {\bf 5} 633

\bibitem{dantan12} Dantan A, Albert M and Drewsen M 2012 {\it Phys. Rev. A} {\bf 85} 013840



\bibitem{muller97} M\"{u}ller G, M\"{u}ller M, Wicht A, Rinkleff R-H and Danzmann K 1997 {\it Phys. Rev. A} {\bf 56} 2385

\bibitem{hernandez07} Hernandez G, Zhang J and Zhu Y 2007 {\it Phys. Rev. A} {\bf 65} 053814

\bibitem{wu08} Wu H, Gea-Banacloche J and Xiao M 2008 {\it Phys. Rev. Lett.} {\bf 100} 173602

\bibitem{mucke10} M\"{u}cke M, Figueroa E, Borchmann J, Hann C, Murr K, Ritter S, Villas-Boas C J and Rempe G 2010 {\it Nature} {\bf 465} 755

\bibitem{kampschulte10} Kampschulte T, Alt W, Brakhane S, Ekstein M, Reimann R, Widera A and Meschede D 2010 {\it Phys. Rev. Lett.} {\bf 105} 155603

\bibitem{laupetre11} Laupr\^{e}tre T, Proux C, Ghosh R, Schwartz S, Goldfarb F and Bretenaker F 2011 {\it Opt. Lett.} {\bf 36} 1551

\bibitem{specht11} Specht H P, N\"{o}lleke C, Reiserer A, Uphoff M, Figueroa E, Ritter S and Rempe G 2011 {\it Nature} {\bf 473} 190


\bibitem{nagorny03} Nagorny B, Els\"{a}sser T and Hemmerich A 2003 {\it Phys. Rev. Lett.} {\bf 91} 153003

\bibitem{brennecke07} Brennecke F, Donner T, Ritter S, Bourdel T, K\"{o}hl M and Esslinger T 2007 {\it Nature} {\bf 450} 268

\bibitem{colombe07} Colombe Y, Steinmetz T, Dubois G, Linke F, Hunger D and Reichel J 2007 {\it Nature} {\bf 450} 272


\bibitem{hornekaer01} Hornek{\ae}r L, Kj{\ae}rgaard N, Thommesen A M and Drewsen M 2001 {\it Phys. Rev. Lett.} {\bf 86} 1994

\bibitem{herskind08} Herskind P F, Dantan A, Langkilde-Lauesen M B, Mortensen A, S{\o}rensen J L and Drewsen M 2008 {\it Appl. Phys. B: Lasers Opt.} {\bf 93} 373

\bibitem{mortensen07} Mortensen A, Nielsen E, Matthey T and Drewsen  M 2007 J. Phys. B: At. Mol. Opt. Phys. {\bf 40}, F223-F229

\bibitem{mortensen04} Mortensen A, Lindballe J J T, Jensen I S, Staanum P, Voigt D and Drewsen M 2004 Phys. Rev. A {\bf 69}, 042502


\end{thebibliography}
\end{document}